\documentclass[prb,twocolumn,showpacs,floatfix,superscriptaddress]{revtex4}
\usepackage{graphicx,epsfig,amssymb}
\usepackage{color}

\begin{document}

\title{Electronic structures and optical spectra of thin anatase TiO$_2$
nanowires through hybrid density functional and quasiparticle calculations}

\author{Hatice \"{U}nal}
\affiliation{Department of Physics, Bal{\i}kesir University,
Bal{\i}kesir 10145, Turkey}

\author{O\u{g}uz G\"{u}lseren}
\affiliation{Department of Physics, Bilkent University,
Ankara 06800, Turkey}

\author{\c{S}inasi Ellialt{\i}o\u{g}lu}
\affiliation{Basic Sciences, TED University, Ankara 06420, Turkey}

\author{Ersen Mete}
\email{emete@balikesir.edu.tr}
\thanks{Corresponding author}
\affiliation{Department of Physics, Bal{\i}kesir University,
Bal{\i}kesir 10145, Turkey}

\date{\today}

\begin{abstract}
The electronic properties of quasi-one-dimensional anatase TiO$_2$
nanostructures, in the form of thin nanowires having (101) and (001) facets,
have been systematically investigated using the standard, hybrid density
functional and quasiparticle calculations. Their visible photoabsorption
characteristics have also been studied at these levels of theories. The thin 
stoichiometric nanowire models are predicted to have larger band gaps
relative to their bulk values. The band gap related features appear to be 
better described with the screened Coulomb hybrid density functional 
method compared to the standard exchange--correlation schemes. 
Depending on the self-consistency in the perturbative GW methods, even 
larger energy corrections have been obtained for the band gaps of both 
(101) and (001) titanium dioxide nanowires.
\end{abstract}


\maketitle

\section{Introduction}

Demand on efficient utilization of solar energy has drawn increasing attention
to reducible metal oxides. The wide-gap semiconductor TiO$_2$ has gained
utter importance in photovoltaics and photocatalysis due to its catalytically
active and reducible surfaces, long standing stability, vast availability,
and nontoxicity.\cite{Diebold} Under UV irradiation TiO$_2$ achieves hydrogen
production from water since the position of the conduction band (CB) well
aligns with the formation energy of hydrogen.\cite{Fujishima1} In addition to
these properties, TiO$_2$ also has excellent charge carrier conduction
features making it one of the best choices as the anode electrode in dye
sensitized solar cells (DSSC).\cite{ORegan,Hangfeldt} Great effort has been
made to extend its UV limited photoresponse to visible region by various
adsorptional, substitutional or interstitial impurities.\cite{Fujishima2,Gratzel,Khan,Chen1}
Along with modification of the electronic structures, in this way, the already
rich photocatalytic properties of titania can be further enhanced.\cite{Zhu,Yin,Celik1}

Among the three polymorphs of TiO$_2$, the anatase phase shows the
highest photocatalytic activity especially with (001) and (101)
surfaces.\cite{Hengerer,Lazzeri,Thomas,Selloni} Although the rutile phase
with (110) bulk termination forms a relatively more stable surface~\cite{Heinrich},
anatase has been reported to be the most stable structure at
nanodimensions.\cite{Naicker,Boercker,Iacomino,Fuertes} Quasi-one-dimensional
nanostructures have large surface-to-volume ratios. In particular, for the
case of titania, this can be benefited in enhancement of efficiencies of
photovoltaic and photocatalytic applications.

Nano-sized materials come into view with preferable and interesting physical
and chemical properties.\cite{Chen2,Cakir1,Cakir2}  
For instance, quasi-one-dimensional
periodic structures facilitate the transport of the charge carriers. Moreover,
relaxation of surface strain during nanowire growth on a semiconductor
substrate naturally avoids lattice mismatch problems observed in the thin film 
case. This allows fabrication of defect-free materials.\cite{Yang}
Single-crystalline anatase TiO$_2$ nanowires were synthesized by
Zhang~\textit{et al.} by using anodic oxidative hydrolysis and hydrothermal
method.\cite{Zhang} Sol-gel coating,\cite{Caruso,Lei} and simple thermal
deposition\cite{Xiang} methods were also successfully used to prepare highly
crystalline anatase nanowires. Jankulovska \textit{et al.} fabricated
well-crystallized TiO$_2$ nanowires about 2 nm in diameter, using chemical bath
deposition at low temperature.\cite{Jankulovska} Recently, Yuan \textit{et al.}
achieved a controlled synthesis of thin-walled anatase nanotube and nanowire
arrays using template-basis hydrolysis.\cite{Yuan} Experimentally prepared
thin nanowires show photoelectrochemical properties different from
nanoparticulate TiO$_2$ electrodes. Especially, an increase in their band
gap energies and photocatalytic oxidation powers was observed, which is
attributed to the quantum-size effect.\cite{Lee,Jankulovska,Yuan,Berger}
Moreover, nanowire systems are capable of showing superior charge carrier
transport features due to their one-dimensional nature.

The band gap-related properties of titania nanostructures have also
been studied by several experiments.\cite{Lee,Jankulovska,Yuan,Berger,Gloter}
Lee \textit{et al.} used UV-vis spectra to demonstrate the band-gap modulation
with particle size (ranging from 3 to 12 nm) in mesoporous TiO$_2$
nanomaterials.\cite{Lee} Yuan \textit{et al.} analyzed the tunability of the
optical absorption edge of TiO$_2$ nanotubes and nanowires with respect to
wall thickness and internal diameter.\cite{Yuan} Similar observations
have been reported by Jankulovska \textit{et al.} for very thin anatase
nanowires.\cite{Jankulovska} Gloter \textit{et al.} studied the energy
bands of titania-based nanotubes with lateral size of $\sim$10 nm using
electron energy-loss spectroscopy (EELS).\cite{Gloter}

On the theoretical side, Szieberth \textit{et al.}, recently, investigated
the atomic and electronic structure of lepidocrocite anatase nanotubes.\cite{Szieberth}
Fuertes \textit{et al.} studied the absorption characteristics of nanostructured
titania by using a self-consistent density functional tight-binding method.\cite{Fuertes}
Tafen and Lewis\cite{Tafen}, and later on, Iacomino \textit{et al.}\cite{Iacomino}
analyzed the effect of size and facet structure on the electronic properties of
anatase TiO$_2$ nanowires within the density functional theory (DFT) approach.
The standard density functional exchange-correlation schemes tend to underestimate
the fundamental band gap of titania by about 1 eV. Moreover, they fall short
in describing defect related gap states.\cite{Celik1} Therefore, proper theoretical
description of the electronic and optical structures of quasi-one-dimensional TiO$_2$
nanomaterials is necessary for a fundamental understanding in terms of pure science 
and for designing more efficient applications in terms of technology.

We employed standard and range separated screened Coulomb hybrid density
functional methods and GW quasiparticle calculations to investigate the
electronic properties, energy corrections and visible absorption profiles
for thin stoichiometric anatase TiO$_2$ nanowire models having (101) and (001)
facets. Our atomistic models represent the smallest possible diameter nanowires.
Therefore, size effects might become apparent and can be discussed
at different flavors of DFT based approaches considered in this work.

\section{Computational Method}

We carried out total energy DFT computations using projector-augmented waves 
(PAW) method~\cite{Kresse1,Blochl,Kresse2} to describe the ionic cores and 
valance electrons with an energy cutoff value of 400 eV for the plane wave 
expansion. Perdew--Burke--Ernzerhof (PBE) functional\cite{Perdew} based on the 
generalized gradient approximation (GGA) has been used to treat nonlocal 
exchange--correlation (XC) effects as implemented in the Vienna ab-initio 
simulation package (VASP).\cite{Kresse1} The Brillouin zone was sampled using 
10$\times$2$\times$2 mesh of \textbf{k}-points.

Inherent shortcoming of the standard DFT due to the lack of proper self-energy 
cancellation between the Hartree and exchange terms as in Hartree--Fock theory, 
causes the well-known band gap underestimation. In particular, strongly 
correlated $3d$ electrons localized on Ti atoms are not properly described. One
of the alternatives to compensate this localization deficiency appears to be 
the screened Coulomb hybrid density functional method, HSE\cite{Heyd1,Heyd2,Paier}, 
which partially incorporates exact Fock exchange and semilocal PBE exchange 
energies for the short range (SR) part as,
\begin{equation}\label{eqn1}
E_{\textbf{\tiny X}}^{\textrm{\tiny HSE}}=
a E_{\textbf{\tiny X}} ^{\textrm{\tiny HF,SR}}(\omega)+
(1-a)E_{\textbf{\tiny X}} ^{\textrm{\tiny PBE,SR}}(\omega)+
E_{\textbf{\tiny X}} ^{\textrm{\tiny PBE,LR}}(\omega)
\end{equation}
where $a$ is the mixing coefficient~\cite{Perdew2} and $\omega$ is the range 
separation parameter.\cite{Heyd1,Heyd2,Paier} The long range (LR) part of 
exchange and full correlation energies are defined by standard PBE~\cite{Perdew} 
functional.

For the description of excitation processes in an interacting many particle 
system, Green's function theory is one of the appropriate methods
through computation of the quasiparticle energies.\cite{Landau,Galitskii}
The quasiparticle (QP) concept makes it possible to describe the system 
through a set of equations, 
\begin{equation}\label{eqn2}
\left(T\!+\!V_{e\textrm{\tiny -}n}\!+\!V_{\textrm{\scriptsize H}}
\!-\!E_{i\textbf{\scriptsize k}}\right)\!\psi_{i\textbf{\scriptsize k}}(\textbf{r})
+\!\int\!\! \Sigma(\textbf{r},\textbf{r}',\!E_{i\textbf{\scriptsize k}}) 
\psi_{i\textbf{\scriptsize k}}(\textbf{r}') d\textbf{r}'= 0
\end{equation}
where $T$ is the kinetic energy operator, $V_{e\textrm{\tiny -}n}$ represents 
the electron--ion interactions, $V_{\textrm{\scriptsize H}}$ is the Hartree 
potential, $E_{i\textbf{\scriptsize k}}$ are the quasiparticle 
energies labeled by state number $i$ and wave vector $\textbf{k}$. The 
self-energy operator $\Sigma$ accounts for the exchange and correlation 
effects and is given by
\begin{equation}\label{eqn3}
\Sigma(\textbf{r},\textbf{r}',\omega)=\frac{i}{2\pi}\int_{-\infty}^{\infty}
e^{i\omega'\delta}G(\textbf{r},\textbf{r}',\omega+\omega')W(\textbf{r},\textbf{r}',\omega')
d\omega'
\end{equation}
where $G$ is the Green's function representing the propagation of a hole or an 
additional particle in the presence of an interacting many particle system, and
$W$ is the dynamically screened Coulomb interaction. The QP energies can be 
determined iteratively by
\begin{equation}\label{eqn4}
E_{i\textbf{\scriptsize k}}^{N+1}=E_{i\textbf{\scriptsize k}}^{N}+Z_{i\textbf{\scriptsize k}}
\textrm{Re}\big[\langle\psi_{i\textbf{\scriptsize k}}|
T+V_{e\textrm{\tiny -}n}+V_{\textrm{\scriptsize H}}+\Sigma(E_{i\textbf{\scriptsize k}})
|\psi_{i\textbf{\scriptsize k}}\rangle\big]
\end{equation}
where $Z_{i\textbf{\scriptsize k}}$ is the normalization factor.\cite{Shishkin} We 
used PBE energy eigenvalues as the starting point and set 
$E^1_{i\textbf{\scriptsize k}}=E^{\textrm{\tiny PBE}}_{i\textbf{\scriptsize k}}$ in 
Eq.~(\ref{eqn4}) to get single shot G$_0$W$_0$~\cite{Hybertsen,Godby} energy 
corrections up to the first-order perturbation theory. In the GW$_0$ case, 
the propagator in Eq. (\ref{eqn3}) is updated after the first iteration while
screened Coulomb term, W, remains fixed. 

\begin{figure*}[t!]
\includegraphics[width=15cm]{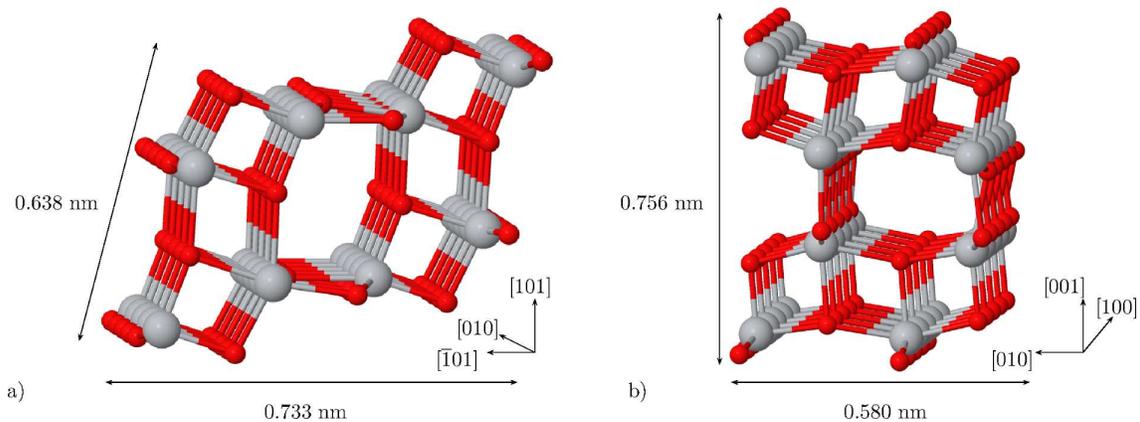}
\caption{Relaxed atomistic structures of the anatase (101) and (001)-nanowire models.\label{fig1}}
\end{figure*}

Shishkin~\textit{et al.}\cite{Shishkin} proposed a self-consistent GW (scGW) 
approach by recasting single-electron theory into the generalized eigenvalue 
problem after linearization around 
some reference energy $E_{i\textbf{\scriptsize k}}^{N}$ :
\begin{eqnarray}\label{eqn5}
&&\mbox{\hspace{-0.7cm}}\nonumber\overbrace{[T+V_{e\textrm{\tiny -}n}+
V_{\textrm{\scriptsize H}}+\Sigma(E_{i\textbf{\scriptsize k}}^N)
+\xi(E_{i\textbf{\scriptsize k}}^N)
E_{i\textbf{\scriptsize k}}^N]}^{\textrm{H}(E_{i\textbf{\scriptsize k}}^N)}
|\psi_{i\textbf{\scriptsize k}}\rangle
\\[1mm]
&&\mbox{\hspace{3.7cm}}=\underbrace{E
(1-\xi(E_{i\textbf{\scriptsize k}}^N))}_{\textrm{S}(E_{i\textbf{\scriptsize k}}^N)}
|\psi_{i\textbf{\scriptsize k}}\rangle
\end{eqnarray}
where H is the non-Hermitian Hamiltonian, S is the overlap operator, and 
$\xi(E_{i\textbf{\scriptsize k}}^N)=\frac{\partial\Sigma(E_{i\textbf{\scriptsize k}}^N)}
{\partial E_{i\textbf{\scriptsize k}}^N}$.
Then, this can be mapped to a simple diagonalization problem, using
the Hermitian parts of H and S matrices, $H$ and $S$, in the DFT basis $\{\phi_i^N\}$, 
\begin{equation}\label{eqn6}
S^{-1/2}HS^{-1/2}U=U\Lambda 
\end{equation}
where $U$ is a unitary matrix and $\Lambda$ is the diagonal eigenvalue 
matrix.\cite{Shishkin} The wave functions are iteratively updated by 
$\phi_i^{N+1}=\sum_i U_{ij}\phi_i^N$ and the corresponding energies
are $E^{N+1}_i=\Lambda_{ii}$. This approximation to the non-Hermitian problem 
in Eq.~(\ref{eqn5}) results in $\sim 1\%$ deviation in band gaps.

Electron--hole interactions can be described by Bethe--Salpeter equation (BSE) 
for the two particle Green's function. In linear-response time-dependent 
density functional theory (TDDFT), the many-body effects are contained in the 
frequency dependent exchange--correlation kernel, 
$f_{\textrm{\small xc}}(\textbf{r}_1,\textbf{r}_2;\omega)$.
Reining\cite{Reining} \textit{et al.} derived a TDDFT XC-kernel from BSE to 
reproduce excitonic effects. Adragna~\textit{et al.}\cite{Adragna} and 
Bruneval~\textit{et al.}\cite{Bruneval} suggested a similar approach to calculate 
the polarizability of a many-body system within the GW framework,
\begin{equation}\label{eqn7}
\chi=[1-\chi_0(v+f_{\textrm{\small xc}})]^{-1}\chi_0 
\end{equation}
where $\chi_0$ is the independent QP polarizability and $v$ is the bare Coulomb 
kernel. We have included electron--hole interactions in our scGW calculations 
using Eq.~(\ref{eqn7}) as implemented in VASP.\cite{Shishkin}

The absorption spectra can be obtained by considering the 
transitions from occupied to unoccupied states within 
the first Brillouin zone. The imaginary part of the dielectric 
function $\varepsilon_2(\omega)$ is given by 
the summation,

\begin{eqnarray} 
&& \hspace{-1cm} \varepsilon^{(2)}_{\alpha \beta}(\omega)=\frac{4 \pi^2 e^2}{\Omega}
\lim_{q\to 0}\frac{1}{q^2}\sum_{c,v,\mathbf{k}}
2\textsl{w}_{\mathbf{k}}\delta(\epsilon_{c\mathbf{k}}-\epsilon_{v\mathbf{k}}-\omega) 
\nonumber \\
&& \hspace{2.4cm}\times\langle u_{c\mathbf{k}+\mathbf{e}_\alpha q} 
\vert u_{v\mathbf{k}} \rangle\langle u_{c\mathbf{k}+\mathbf{e}_\beta q} 
\vert u_{v\mathbf{k}} \rangle^*
\end{eqnarray}

\noindent where the indices $c$ and $v$ indicate empty and 
filled states respectively, $u_{c\mathbf{k}}$ are the cell periodic 
part of the orbitals and $\textsl{w}_{\mathbf{k}}$ are the weight factors at each 
\textbf{k}-point.\cite{Gajdos}

\section{Results \& Discussion}

\begin{figure*}[t!]
\includegraphics[width=16.5cm]{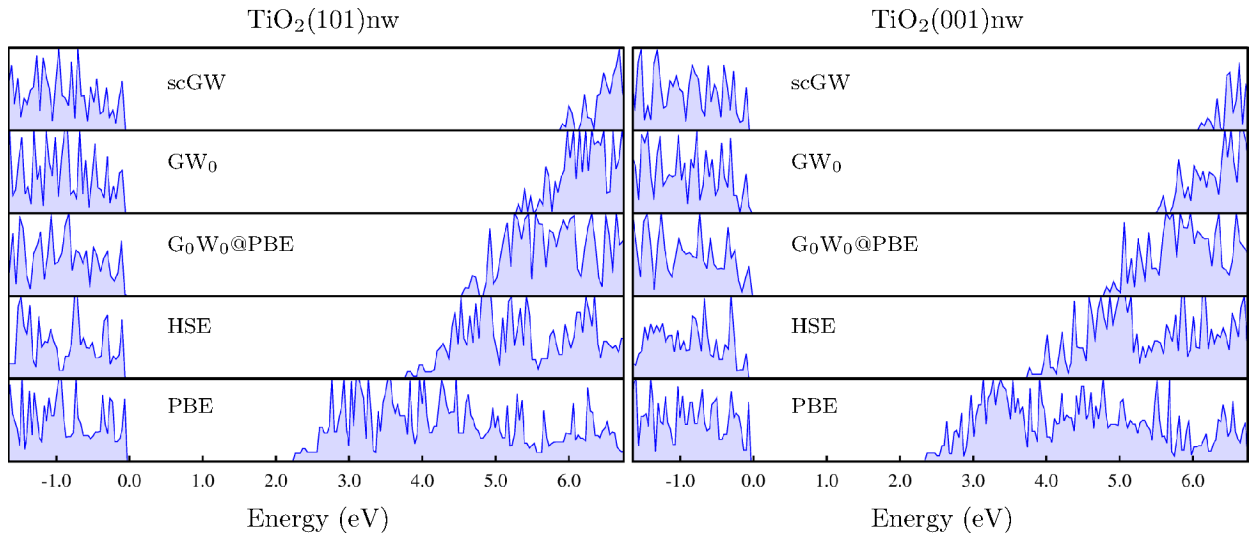}
\caption{Densities of states (DOS) plots of bare anatase (101)-nanowire and 
(001)-nanowire calculated using PBE, HSE functionals within DFT and G$_0$W$_0$, 
GW$_0$, scGW methods within many-body perturbation theory (MBPT) starting from 
PBE initial wavefunctions. The origin of the energy axis is set at just above 
the VBM.\label{fig2}}
\end{figure*} 

The minimum-energy band gap of bulk anatase TiO$_2$ with the standard PBE XC 
functional is found to be 2.03 eV indirect between $\Gamma$ and a point close to 
X while the direct gap at $\Gamma$ is 2.35 eV. These are inconsistent 
with the experimental results (3.2$-$3.4 eV).\cite{Tang,Kavan} The local density 
approximation (LDA/GGA) tends to distribute charge based on the properties of 
an homogeneous electron gas. In the case of TiO$_2$, this leads to an 
unsatisfactory description of localized $3d$ states of Ti. A Hubbard U term 
can be added only for the $d$-space in order to supplement repulsive 
correlation effects between the $d$-electrons. We performed a simple PBE+U 
calculation with U = 5 and get a band gap of 2.56 eV for the bulk anatase. Larger 
values of U increase this value but distort the lattice structure unacceptably. 
The range-separated hybrid DFT approach has a potential to improve energy gap 
related properties by incorporating HF exchange interaction. We previously 
found an indirect gap value of 3.20 eV using the HSE method with a mixing 
factor of $a=0.22$.\cite{Celik2} Another alternative is to use perturbation 
theory to get quasiparticle energy shifts. In recent studies on TiO$_2$, 
Chiodo~\textit{et al.}\cite{Chiodo}, Landmann~\textit{et al.}\cite{Landmann} 
and Kang~\textit{et al.}\cite{Kang} calculated the indirect electronic gap as 
3.83 eV, 3.73 eV and 3.56 eV, respectively, at the single shot G$_0$W$_0$ 
level. Noticeable disagreement with the experiments is due to the choice of the 
starting point from the inaccurate DFT description of Ti $3d$ states. 
Patrick~\textit{et al.}\cite{Patrick} reported a gap of 3.3 eV 
by performing G$_0$W$_0$ calculation starting from DFT+U band structure while 
single shot GW on top of DFT wave functions gave a value of 3.7 eV. This 
approach still depends on the empirical U parameter even though it is 
computationally less demanding. For a parameter-free theory, one needs a 
self-consistent GW procedure. But self-consistency solely can not give desired 
accuracy without including electron--hole interactions. For this reason, we 
performed scGW calculations including vertex corrections\cite{Adragna,Bruneval} 
and calculated an electronic gap of 3.30 eV for the bulk anatase in good 
agreement with the experiments.\cite{Tang,Kavan}

For the quasi-two-dimensional cases, the optical spectra of the anatase surface 
is essentially similar to the absorption and photoluminescence (PL) data of the 
bulk.\cite{Giorgi} Giorgi \textit{et al.}\cite{Giorgi} identified the first 
direct exciton at $\sim$3.2 eV on the anatase (001)-(1$\times$1) surface from
the QP calculations. At the nanoscale, the reduction of material sizes below 
the exciton radius gives rise to an increase of the band gap as a quantum 
confinement effect. The exciton radii for titania were estimated in between
0.75 nm and 1.9 nm.\cite{Kormann,Serpone} The blue shift of the band gap 
becomes dominant for materials with cross section sizes fitting in this range.

In order to discuss the electronic structure and possible size effect at 
different level of density functional theories, thin nanowire models were built 
from the anatase form of TiO$_2$ having (001) and (101) facets. They will be 
referred as nw(001) and nw(101), respectively. We preserved the stoichiometry in 
building atomistic models, and did not saturate any of the dangling bonds 
exposed on the facets. Nanowire structures have been represented in a 
tetragonal supercell geometry using periodic boundary conditions, PBC. While 
the PBC along the nanowire axis leads to infinitely long wire, to prevent 
interaction between adjacent isolated wires, a large spacing of at least 
20 {\AA} perpendicular to the axis has been introduced. Initial geometries have 
been fully optimized based on the minimization of the Hellman--Feynmann forces 
on each of the atoms to be less than 0.01 eV/{\AA}. The relaxed atomistic 
structures of the anatase nw(101) and nw(001) models as shown in 
Fig.~\ref{fig1} do not show any major reconstruction from their initial 
configurations cleaved from bulk structures. The Ti--O bond lengths on the facets 
get slightly larger than the bulk value of 1.95 {\AA}. This deviation is much 
less inside the nanowire maintaining  the anatase form for these isolated 
free-standing 1D thin nanostructures. Relaxation of surface atoms passivates 
possible surface states to appear in the band gap (see Fig~\ref{fig2}).

\begin{table}[tbh]
\caption{Calculated band gaps (in eV) of TiO$_2$ nanowires\label{table1}}
\begin{ruledtabular}
\begin{tabular}{cccccccccc}
Nanowire&PBE&HSE06&G$_0$W$_0$@PBE&GW$_0$&scGW\\[1mm] \hline
{(101)}&2.51&4.01&4.88&5.60&6.05\\ \hline
{(001)}&2.69&4.06&5.15&5.79&6.25
\end{tabular}
\end{ruledtabular}
\end{table}

For the TiO$_2$ nanotubes with internal diameters in the range $2.5-5$ nm, 
Bavykin \textit{et al.}\cite{Bavykin1,Bavykin2} estimated an optical gap of 
3.87 eV from their absorption and PL studies. Yuan \textit{et al.} reported a 
significant blue shift of the optical absorption edge as the wall thickness of 
anatase nanotubes decrease from 45 to 10 nm.\cite{Yuan} Similarly, the energy 
gap was reported to be 3.84 eV for 2D titanate nanosheets~\cite{Sakai}, 
and to be 3.75 eV for thin anatase TiO$_2$ films.\cite{Park} These are 
significantly larger than the bulk value of 3.2 eV. 

We present the band gap values of  thin anatase nw(101) and nw(001) structures 
calculated with various levels of theory in Table~\ref{table1}. Although still 
underestimated, the standard PBE functional gives band gaps for these 1D systems 
larger than the bulk value of 2.03 eV.  Admixing partial exact exchange energy 
through a screened Coulomb interaction, HSE method predicts the gaps as 
4.01 eV and 4.06 eV for nw(101) and nw(001), respectively. Therefore, hybrid 
HSE functional largely improves over PBE results. Size effect for nw(001) and 
nw(101) having diameters $\sim$0.75 nm becomes remarkable at the hybrid DFT 
level. Even though hybrid DFT is not designed to describe absorption processes, 
the positions of lowest lying absorption peaks can reasonably be estimated by 
these methods.\cite{Celik1,Celik2} 

One of the methods to describe excitations is the time dependent density 
functional theory (TDDFT). Meng \textit{et al.} used TDDFT method 
on a hydrogenated nanowire segment having anatase (101) facet as a finite 
system. Although, their nanowire segment is thicker than our model structures, 
the optical spectrum of the bare nanowire using TDDFT shows an increase in the 
band gap relative to the bulk value. Unexpectedly, 
Fuertes \textit{et al.}\cite{Fuertes} predicted an energy gap of 2.92 eV for an 
anatase cluster composed of 34 TiO$_2$ units using a time-dependent density 
functional tight-binding method. However, they mention possible involvement 
of surface states narrowing the gap. 

\begin{figure*}[htb]
\includegraphics[width=15.4cm]{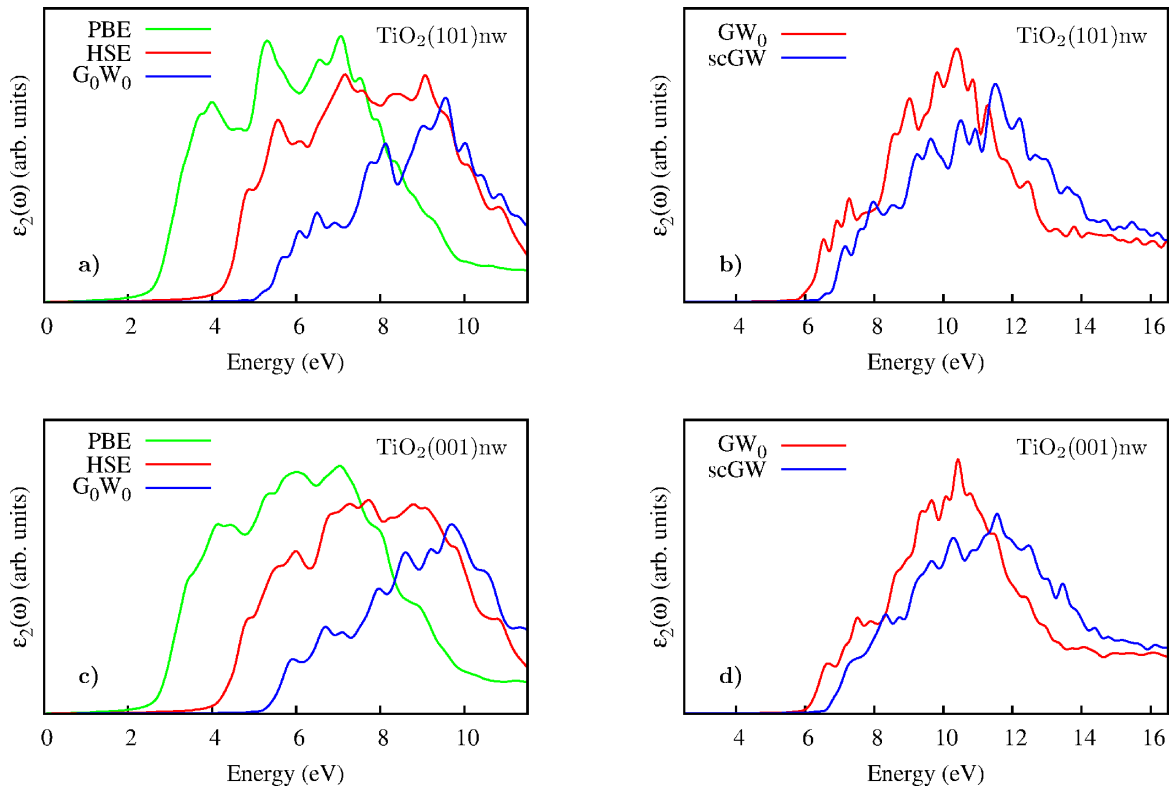}
\caption{Calculated absorption spectra for the anatase TiO$_2$ nw(101) and nw(001) 
nanowire models calculated with density functional PBE, HSE, G$_0$W$_0$, GW$_0$, 
and scGW methods.\label{fig3}}
\end{figure*}

Several experimental studies observed a blue shift of the optical gap as TiO$_2$ 
nanomaterial sizes decrease.\cite{Anpo,Joselevich,Serpone,Kumar} The liable 
quantum confinement effect is reported at different size regimes. For instance, 
Kumar~\textit{et al.}\cite{Kumar} reported a linear decrease in the energy gap 
from 3.83 eV to 3.70 eV with an increase in the fiber diameter from 60 nm to 
150 nm. Anpo~\textit{et al.}\cite{Anpo} observed the size effect for particle 
sizes of several tens of nanometers while Serpone~\textit{et al.}\cite{Serpone} 
identified it for nanometer sized colloidal anatase particles. 
Lee~\textit{et al.}\cite{Lee} estimated an inverse proportionality of the band gap 
to the nano particle size. For anatase, their prediction gets as large as 4 eV at 
a particle size of 2 nm.\cite{Lee} 

For a (0,n) lepidocrocite-type TiO$_2$ nanotube with a diameter of 1.81 nm, 
Szieberth~\textit{et al.} calculated a band gap of 5.64 eV using a density 
functional theory--tight binding (DFT-TB) method.\cite{Szieberth} In a 
previous GW study, Mowbray~\textit{et al.} calculated the quasiparticle gap 
of a (4,4) TiO$_2$ nanotube having a diameter of 0.8 nm to be about 
7 eV.\cite{Mowbray} This QP gap value reflects an overestimation associated 
with the lack of self-consistency and excitonic effects in their GW 
calculations. Therefore, it was suggested as an upper bound for the optical 
gap. For the thin anatase (101) and (001) nanowires, $~\sim$0.75 nm in 
diameter, our G$_0$W$_0$-predicted QP gaps are 4.88 eV and 5.15 eV, 
respectively. The blue shift of the gap is attributed to the quantum 
confinement effect which is strong in this size regime. In this sense, 
for the bare nanowires, our QP results can be considered more reliable 
relative to hybrid DFT methods where a portion of the exact exchange is mixed 
with the PBE exchange. 

Along with the calculated energy gaps, similar conclusions can be drawn from 
the density of states, DOS, presented in Fig.~\ref{fig2}. The valance band (VB) 
edge showing O $2p$ character slightly changes depending on the XC functional 
used or on the level of quasiparticle calculation performed. On the other hand, 
the conduction band (CB) edge which is mainly formed from Ti $3d$ states shifts 
and sets the value of the electronic band gap. 

In GW$_0$ calculations, the self-consistency is imposed on the single particle 
propagator giving rise to larger energy corrections relative to those of 
G$_0$W$_0$. So, the QP gaps become to 5.60 eV for nw(101) and 5.79 eV for 
nw(001). The self-consistency in both the single particle propagator and 
the dynamical screening tends to shift the unoccupied Ti $3d$ states up to 
much higher energies. In our scGW calculations including electron--hole 
interactions, we obtained the QP gaps as 6.05 eV and 6.25 eV for the thin (101) 
and (001) nanowires, respectively. A trend of increasing energy correction with 
increasing level of theory is seen. A direct comparison of QP or hybrid DFT 
results with the experimental data is generally not straightforward due to 
possible involvement of stress, impurity or defect related states. Even so,
our scGW calculations estimate the QP gaps in good agreement with previous
experimental\cite{Lee} and theoretical\cite{Mowbray,Szieberth} findings.

For the discussion of the absorption spectra, the imaginary part of the 
dielectric function for anatase (101) and (001) nanowires are depicted in 
Fig.~\ref{fig3}. For both of the nanowires, in all cases, the absorption starts 
around the conduction band edge energies consistent with the calculated band 
gaps. The VB maximum is dominantly populated with O $2p$ electrons. The CB 
minimum is characterized by Ti $3d$ $t_{2g}$-states. Hence, one can conclude 
that the first peak mainly contributed by the transitions from the states at 
the VB top to the states at the CB edge. Therefore, these transitions are 
dipole-allowed and are suitable for photocatalytic applications. The 
scGW-calculated optical spectra shows that the photo-response of defect-free 
anatase TiO$_2$ nanowires significantly blue shift up in the UV region for 
nanowire radii within the quantum confinement regime. In other words, as we 
employed more accurate density functional based theories starting  from the 
standard PBE up to scGW including excitonic effects, we have obtained a trend 
of increasing blue shifts in the band gaps of anatase nanowires with 
diameters around 1 nm. Experimental observation of such a large quantum 
size effect might be concealed by possible presence of stress, impurity or 
defect related gap states. 

\section{Conclusions}

In summary, the electronic band gap and absorption properties of thin TiO$_2$
nanowires having (101) and (001) facets have been investigated at the levels of
exact exchange mixed hybrid DFT and quasiparticle calculations with various
self-consistency schemes. When the periodicity is reduced to one dimension
as in the nanowire model structures, the small diameters result in larger electronic
band gaps. Therefore, the dimensionality of the nano materials plays a critical
role in the photoresponse of titania. Moreover, dye adsorbates or transition metal
dopants will be crucially important to functionalize these semiconductor metal
oxides under visible light illumination at the nano-scale. Such impurities also
greatly influence efficiencies for photovoltaic and photocatalytic applications.

Range separated hybrid functionals incorporating exact exchange with $1/r$
Coulomb tail improve the electronic description of TiO$_2$ nanowires.
Although they are not intended to get the excited state properties,
photo absorption characteristics  are also healed relative to traditional
semi-local exchange-correlation schemes due partly to the shift of
unoccupied states to higher energies.  In order to get proper description
of excited state properties one has to include electronic screening effects.
This can be achieved by non-empirically range separated hybrid
approaches or many body perturbative methods to calculate self-energy
contributions. Higher levels of density functional theory increases accuracy
at a computational cost.  Consequently, a practical and reliable determination 
of size dependence of  excitation gaps in TiO$_2$ nano-materials are still 
desirable.

\begin{acknowledgments}
This work was supported by T\"{U}B\.{I}TAK, The Scientific and Technological
Research Council of Turkey (Grant \#110T394). Computational resources were
provided by ULAKB\.{I}M, Turkish Academic Network and Information Center.
\end{acknowledgments}

\end{document}